\documentclass{pasj00}

\begin{document}
\SetRunningHead{Iono et al.}{High-z CO Observations at the NRO 45m}

\title{Initial Results from the Nobeyama Molecular Gas Observations of Distant Bright Galaxies}

\author{Daisuke \textsc{Iono}\altaffilmark{1,2}, 
Bunyo \textsc{Hatsukade}\altaffilmark{3}, 
Kotaro \textsc{Kohno}\altaffilmark{4,5}, 
Ryohei \textsc{Kawabe}\altaffilmark{1}, 
Soh \textsc{Ikarashi}\altaffilmark{5}, 
Kohei \textsc{Ichikawa}\altaffilmark{3},
Tadayuki \textsc{Kodama}\altaffilmark{2,6}, 
Kentaro \textsc{Motohara}\altaffilmark{5}, 
Taku \textsc{Nakajima}\altaffilmark{1}, 
Koichiro \textsc{Nakanishi}\altaffilmark{2,6,7}, 
Kouji \textsc{Ohta}\altaffilmark{3}, 
Kazuaki \textsc{Ota}\altaffilmark{3}, 
Toshiki \textsc{Saito}\altaffilmark{8}, 
Kenta \textsc{Suzuki}\altaffilmark{5}, 
Kenichi \textsc{Tadaki}\altaffilmark{8}, 
Yoichi \textsc{Tamura}\altaffilmark{5}, 
Junko \textsc{Ueda}\altaffilmark{7}, 
Hideki \textsc{Umehata}\altaffilmark{5}, 
Kiyoto \textsc{Yabe}\altaffilmark{3},
Tessei \textsc{Yoshida}\altaffilmark{3},
Suraphong \textsc{Yuma}\altaffilmark{3}, 
Nario \textsc{Kuno}\altaffilmark{1,2}, 
Shuro \textsc{Takano}\altaffilmark{1,2}, 
Hiroyuki \textsc{Iwashita}\altaffilmark{1}, 
Kazuyuki \textsc{Handa}\altaffilmark{1}, 
Aya \textsc{Higuchi}\altaffilmark{6,7}, 
Akihiko \textsc{Hirota}\altaffilmark{1}, 
Shinichi \textsc{Ishikawa}\altaffilmark{1}, 
Kimihiro \textsc{Kimura}\altaffilmark{9}, 
Jun \textsc{Maekawa}\altaffilmark{1}, 
Hiroshi \textsc{Mikoshiba}\altaffilmark{1}, 
Chieko \textsc{Miyazawa}\altaffilmark{1}, 
Kazuhiko \textsc{Miyazawa}\altaffilmark{1}, 
Kazuyuki \textsc{Muraoka}\altaffilmark{9},
Hideo \textsc{Ogawa}\altaffilmark{9},  
Sachiko \textsc{Onodera}\altaffilmark{1}, 
Yasufumi \textsc{Saito}\altaffilmark{1}, 
Takeshi \textsc{Sakai}\altaffilmark{5}, 
Shigeru \textsc{Takahashi}\altaffilmark{1},
Min S. \textsc{Yun}\altaffilmark{10}
} %
\altaffiltext{1}{Nobeyama Radio Observatory, 
National Astronomical Observatory of Japan, 
Minamimaki, Nagano, 384-1305}
\altaffiltext{2}{The Graduate University for Advanced Studies 
(SOKENDAI), 2-21-1 Osawa, Mitaka, Tokyo 181-0015}
\altaffiltext{3}{Department of Astronomy, Kyoto University,
Kyoto 606-8502}
\altaffiltext{4}{Research Center for the Early Universe, 
School of Science, University of Tokyo, 7-3-1 Hongo, Bunkyo, Tokyo 113-0033, Japan}
\altaffiltext{5}{Institute of Astronomy, 
The University of Tokyo, 2-21-1 Osawa, Mitaka, Tokyo 181-0051}
\altaffiltext{6}{National Astronomical Observatory of Japan, 
2-21-1 Osawa, Mitaka, Tokyo 181-0051}
\altaffiltext{7}{Joint ALMA Observatory, Alonso de Cordova 3107, 
Vitacura, Santiago, Chile}
\altaffiltext{8}{Department of Astronomy, Graduate School of Science,
The University of Tokyo, 7-3-1 Hongo, Bunkyo-ku, Tokyo 113-0033}
\altaffiltext{9}{Department of Physical Science, 
Osaka Prefectural University, Gakuen 1-1, Sakai, Osaka 599-8531}
\altaffiltext{10}{Department of Astronomy, University of Massachusetts,
Amherst, MA 01002} 

%


\maketitle

\begin{abstract}
We present initial results from the CO survey toward high redshift 
galaxies using the Nobeyama 45m telescope.  
Using the new wide bandwidth spectrometer equipped with a two-beam SIS
receiver, we have robust new detections of three high redshift ($z=1.6-3.4$)
submillimeter galaxies (SXDF~1100.001, SDP9, and SDP17),
one tentative detection (SDSS~J160705+533558), 
and one non-detection (COSMOS-AzTEC1). 
The galaxies observed during the commissioning phase are sources with 
known spectroscopic redshifts from previous optical or from wide-band 
submm spectroscopy.  The derived molecular gas mass and line widths 
from Gaussian fits are $\sim 10^{11}$~M$_{\odot}$ and $430$ -- 
$530$~km~s$^{-1}$, which 
are consistent with previous CO observations of distant 
submm galaxies and quasars.  
The spectrometer that allows a maximum  of 32~GHz 
instantaneous bandwidth will provide new science capabilities at the 
Nobeyama 45m telescope, allowing us to determine  redshifts
of bright submm selected galaxies without any prior redshift information. 
\end{abstract}

\section{Introduction}

Single dish submm telescopes equipped with wide bandwidth bolometer cameras 
have discovered a large population of galaxies in the distant universe 
that are extremely bright in mm/submm wavelengths (submillimeter galaxies (SMGs); e.g. \cite{smail97,hughes98,greve08,scott10,hatsukade10}).  
The bright ($\gtrsim 1$~mJy) SMG population typically shows evidence of 
enhanced massive star formation 
activity (SFR $\gtrsim 500$~M$_{\odot}$~yr$^{-1}$) with 
rapid gas consumption timescales ($\sim40$~Myr: \cite{greve05}) and complex 
kinematics \citep{tacconi08}, which are all consistent with 
the properties expected for gas-rich major mergers at high redshifts.
Results from these early observations are also 
consistent with the scenario where
present-day massive galaxies acquire the bulk of their stellar 
mass at early epochs.

SMGs contribute 10 - 20\% to the 
cosmic star formation at $z \sim 2-3$ \citep{hatsukade10}, but the 
lack of precise redshifts have been the significant source of uncertainty.  
The redshift determination has been further hampered by the faintness 
of the optical counterparts 
even for those that are identified through 
direct submm interferometric imaging 
(e.g.; \cite{iono06,younger07}).  
Uncertainties in the redshifts also 
affect the derivation of the clustering strength 
(Hatsukade et al., in prep) 
and the dark halo mass, which are both important parameters in the 
framework of the cold dark matter (CDM) cosmology.



The most direct way to 
determine the redshifts of SMGs is to conduct wide-band 
spectroscopic observations 
in the radio/mm/submm wavelengths through detections of the 
carbon monoxide (CO) emission.  Recent observations using 
single dish telescopes 
have shown this to be effective 
(i.e. GBT-Zpectrometer, IRAM~30m-EMIR, CSO-Zspec, LMT-RSR; 
\cite{erickson07,weiss09,harris10,lupu10,carter12}).  
Significant detection of two consecutive CO rotational transitions is
needed for a secure determination of the redshift. 
In addition, these observations 
will allow us to quantify the molecular gas mass and its dynamical status
through the CO line width.
This will further allow us to derive the gas mass 
fraction, star formation efficiencies, and the gas-to-dust mass ratio, 
all of which are important quantities related to their physical conditions
and possibly their future evolution. 

The 45m telescope at Nobeyama has undergone a major upgrade, with an
installation of a new two-beam receiver, an intermediate frequency (IF) 
transmission system,
an Analogue-to-Digital Converter (ADC) with a sampling rate of 4~GHz, and 
a new 32~GHz wide spectrometer.
The final goal of this project is to obtain redshifts of the bright submm 
sources blindly from CO observations that are near or beyond the 
peak of the cosmic star formation.  In order to 
assess the capabilities of the newly upgraded 45m telescope,
we have targeted five sources that are bright in submm continuum 
and have spectroscopic redshifts, determined either via 
conventional optical/infrared observations, or via CO. 
The sources we select in this pilot study are, 
SXDF~1100.001 (Orochi) \citep{ikarashi11}, SDP9, 
and SDP17 \citep{negrello10,lupu10}, 
SDSS~J160705+533558 \citep{clements09}, and COSMOS-AzTEC1 \citep{smolcic11}.
These sources have apparent infrared
luminosities of $\gtrsim 10^{13}$~L$_{\odot}$, and are promising sources for 
CO detection.  
We adopt H$_0$ = 71~km~s$^{-1}$~Mpc$^{-1}$, 
$\Omega_{\rm M}$ = 0.27, $\Omega_{\Lambda}$ = 0.73 for all of the 
analysis throughout this paper.




\section{Instrumental Setup and Observations}


The radio signal collected by the 45m telescope dish is received by 
the new two-beam, two-polarization, sideband separating 
receiver (TZ receiver) \citep{nakajima12}.  
The two beams are separated by $\sim45\arcsec$ in the sky.  
The sky signal is separated into two linear polarizations 
by a waveguide-type ortho-mode transducer, and an IF 
quadrature hybrid is then used to separate the two sidebands.
The local oscillator frequency tuning rage is 80 -- 115~GHz,
with an IF bandwidth of 4 -- 8~GHz.  
A future expansion to 4 -- 12~GHz is planned.

The IF signal is 
amplified and 
down-converted to 2 -- 4~GHz where the signal is digitized to 3-bits 
at the ADC. The digital signal is then 
transferred through optical fiber cables to the digital back-end.  
The digital back-end is an exact copy of the 
Atacama Compact Array (ACA) \citep{iguchi09} 
FX-type correlator \citep{kamazaki11}.
The 32~GHz total bandwidth is divided into 16 arrays, each with 2~GHz of
bandwidth.  When used at its maximum capability, it will accommodate 
the full 4 -- 8~GHz IF signal from the TZ receiver.

Observations were carried out during the 
2011/2012 winter observing season.  The single sideband 
system temperature varied from 170 - 250~K 
\footnote{The receiver temperature 
is expected to improve significantly next year with the installation of 
new components.}, and 
we have tuned the orthogonal polarizations of each beam 
to the same sky frequency to gain $\sim \sqrt 2$ in sensitivity.  
The sideband rejection ratio is typically $>$ 8dB at the center of the IF.
Since the source is always observed in one of the beams 
(16 to $19\arcsec$ at the observed frequencies) of the
TZ receiver, the dominant observing overhead is the 
telescope slewing time 
during switching, and it is $\sim 7$~seconds per cycle.
We have adopted a 5 -- 10 second switching cycle 
to optimize the observing overhead and the bandpass stability.
The chopper-wheel method was used for absolute temperature 
calibration with an accuracy of 20\%.  
The output auto-correlation spectrum, which contains 4096~channels with a 
frequency resolution of 488~kHz, is then calibrated on-line and 
stored in the database.

Data reduction was carried out using the facility data reduction 
software package Newstar.  Since the 
45m pointing accuracy is significantly affected by wind, 
only the data taken with wind velocity less than 5~m~s$^{-1}$
were used.
The pointing
accuracy under these conditions is typically 2--$3\arcsec$ 
in RMS.
In addition, we flagged scans with visually poor baselines by inspecting 
each spectrum by eye. 
The final spectra are generated by 
integrating all of the good scans in the data, 
and a zeroth order baseline is subtracted from each spectrum.  
The average on-source time ranged from 5 -- 11 hours.
The antenna temperature is then converted to main-beam temperature 
by adopting a main-beam efficiency of 0.4, which is the average  
value obtained using the T100 receiver near the observing frequencies.
\footnote{See status report; http:$\slash\slash$ www.nro.nao.ac.jp$\slash$ \~{} nro45mrt
$\slash$ prop $\slash$ status $\slash$ Status\_R11.html}
A typical 1$\sigma$ sensitivity after one hour integration is 
$\sim$5--8~mJy under these observing conditions and instrumental setup.

\section{CO Spectra and Derived Quantities}

\begin{figure*}
  \begin{center}
    \FigureFile(150mm,100mm){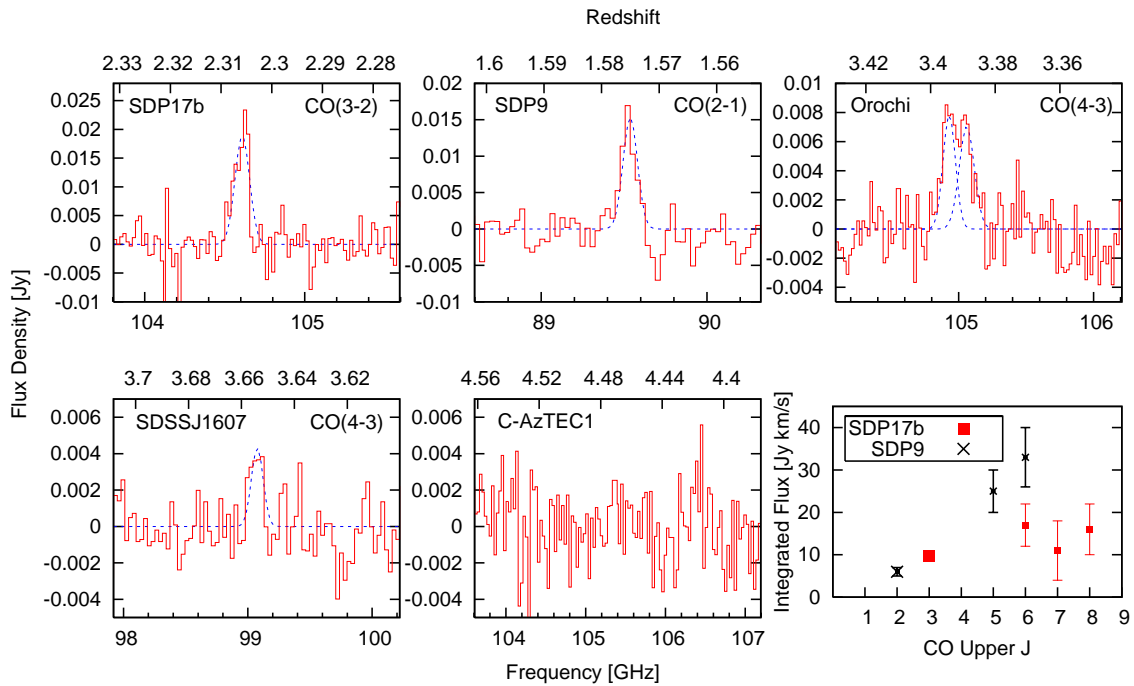}
  \end{center}
  \caption{The CO spectra are binned with 30~MHz spectral resolution for SDP9, SDSS~J1607 and C-AzTEC1, and 20~MHz for SDP17b and Orochi.  Gaussian fits to the detected spectra are shown in dashed blue lines.  
The lower right panel shows the relationship between integrated intensity [in Jy~km~s$^{-1}$] and the CO Upper J transition for the two SDP sources.  The CO~(6--5), CO~(7--6), CO~(8--7) for SDP17b and CO~(6--5), CO~(5--4) for SDP9 are taken from \citet{lupu10}}\label{fig1}
\end{figure*}

The CO spectra are shown in 
Figure~\ref{fig1} and the derived properties  
are summarized in Table~\ref{tab1}.

\subsection{SDP17b \& SDP9}
SDP~17b was originally discovered in the Herschel-Astrophysical Terahertz 
Large Area Survey (H-ATLAS) \citep{eales10} in the Science 
Demonstration Phase (SDP) as a population of submillimeter-bright 
strongly lensed galaxies \citep{negrello10}.  This source was 
subsequently followed up with Z-Spec on board the Caltech Submillimeter
Observatory, detecteding CO~(8--7), CO~(7--6), CO~(6--5) emission
in the 200 -- 300 GHz band \citep{lupu10}.  The derived redshift is
$z=2.308$.  In addition, \citet{omont11} have used the  
Plateau de Bure Interferometer (PdBI) to detect significant H$_2$O emission.

The CO~(3--2) emission from SDP17b is the brightest among all lines detected
in this pilot study.  
The 45m telescope observes a lower rotational transition of CO 
than those detected 
using Z-spec, allowing us to investigate the CO excitation conditions.  
While the uncertainties in the high-J CO line 
fluxes are relatively high, 
the CO excitation (Figure~\ref{fig1}) appears low and may peak near the 
mid-J lines detected at Z-Spec.  This evidence for relatively low 
CO excitation is consistent with other SMGs studied in multi-line CO
\citep{weiss07}. 



SDP9 is another source detected in the H-ATLAS.  The CO~(6--5) and CO~(5--4)
emission have been detected with Z-Spec \citep{lupu10}, yielding a 
redshift of $z=1.578$.  We detected the CO~(2--1) line from SDP9 
with 8-sigma confidence.  
The derived redshift is 0.003 smaller than the Z-Spec redshift.
In contrast to SDP17b, the CO excitation appears high and does not show 
significant signs of decrement at mid-J CO lines.  Detections of 
higher-J CO lines are necessary to understand the excitation conditions 
of this SMG better.

\begin{table*}
\begin{center}
  \caption{Observational Properties and Derived Parameters}\label{tab1}
  \begin{tabular}{lccccccc}
  \hline           
  \multicolumn{1}{c}{Source}  & Transition & RMS~[mJy]\footnotemark[1] & $z$ & FWHM~[km~s$^{-1}$] & $S_{CO}~\Delta v$ [Jy~km~s$^{-1}$]& M$_{\rm H_2}$ [M$_{\odot}$]\footnotemark[2] \\ 
  \hline
  SDP17b    & CO~(3--2) & 4.7 & $2.306 \pm 0.001$ & $440 \pm 30$ & $9.6 \pm 1.9$  & $(4.3 \pm 0.9) \times 10^{11}$\\
  SDP9      & CO~(2--1) & 2.2 & $1.575 \pm 0.001$ & $480 \pm 30$ & $6.1 \pm 1.2$ & $(1.9 \pm 0.4) \times 10^{11}$\\
  Orochi\footnotemark[3]    & CO~(4--3) & 1.9 & $3.394 \pm 0.003$ & $430 \pm 40$ & $5.8 \pm 1.2$  & $(3.4 \pm 0.7) \times 10^{11}$\\
  & & & $3.388 \pm 0.004$ & $530 \pm 40$ & \\
  SDSSJ1607 & CO~(4--3) & 1.7 & $3.653 \pm 0.001$ & $480 \pm 70$ & $1.5 \pm 0.4$ & $(1.1 \pm 0.2) \times 10^{11}$\\
  C-AzTEC1 & -- & 2.0 & -- & -- & $<2.4$ & $< 1.9 \times 10^{11}$  \\
  \hline
  \multicolumn{4}{@{}l@{}}{\hbox to 0pt{\parbox{180mm}{\footnotesize
       \footnotemark[1] The root mean squared of the line free region of the spectrum using the channel binning shown in Figure~1. \par\noindent
       \footnotemark[2] Using CO line ratios in \citet{bothwell12} and 
       the CO-to-H$_2$ conversion factor of 0.8 [M$_{\odot}$~(K~km~s$^{-1}$~pc$^2$)$^{-1}$].  
       \par\noindent
       \footnotemark[3] The redshift and FWHM of Orochi are derived using a two Gaussian fit, but the integrated intensity and mass are for the entire spectrum.
       \par\noindent
     }\hss}}
\end{tabular}
\end{center}
\end{table*}

\subsection{SXDF1100.001~(Orochi/HXMM02)}

SXDF1100.001　(Orochi) 
is a L$_{\rm IR} \sim 10^{13}$~L$_{\odot}$ SMG discovered 
using  
the AzTEC camera \citep{wilson08} on the Atacama 
Submillimeter Telescope Experiment (ASTE) \citep{ezawa04,ezawa08}.  Its 1.1~mm 
flux density is 37~mJy, making this the brightest SMG found in the 
Subaru/XMM-Newton Deep Field (SXDF) and 
may be gravitationally lensed by a foreground galaxy \citep{ikarashi11}.  
This source is detected by Herschel Multi-tiered Extragalactic Survey 
(HerMES), also known as HXMM02, and found to be very bright in 
Herschel/SPIRE bands \citep{wardlow12}.
The photometric redshift was estimated to be $z = 3.4^{+0.7}_{-0.5}$ using 
submm to radio continuum fluxes from SMA, Z-Spec, CARMA, and 
VLA \citep{ikarashi11}. Subsequent CO~(3--2), (4--3), and (5--4) 
detections with CARMA and PdBI confirm that HXMM02 is at 
$z = 3.395$ \citep{wardlow12}. The CO~(1--0) is also measured 
using GBT (Inoue et al. 2012, in prep.). 



The CO~(4--3) line of Orochi is best fit with a two component Gaussian. 
The double peak spectrum may suggest  
a merging galaxy, or a presence of systematic rotation of a 
large gas concentration.  Such double peak profiles are often seen in past
detections toward SMGs \citep{greve05}
and in BzK selected galaxies \citep{daddi10}.  


\subsection{SDSS~J160705+533558}

SDSS~J160705+533558~(SDSS~J1607) is a submm bright quasar at $z=3.653$ 
with an infrared luminosity exceeding $10^{14}$~L$_{\odot}$~\citep{clements09}.
Followup $1\arcsec$ SMA observation has found a resolved 
dust component with the peak located $\sim2\arcsec$ north 
of the optical quasar.  The large submm flux is equivalent to a dusty SFR of  
3000 -- 8000 M$_{\odot}$~yr$^{-1}$.  The submm emission appears to be resolved 
in the north-south direction, and \citet{clements09} suggest either 
a merging galaxy or an interaction with an 
AGN jet for the unusual submm morphology.

We tentatively detect the CO~(4--3) line in SDSS~J1607.  
The linewidth and molecular gas mass are both consistent with 
other high redshift quasars  
detected in CO \citep{wang10}, but the 
gas depletion time scale (M$_{\rm H_2}/$SFR) of 14 -- 37~Myr 
is slightly smaller than the average of quasars \citep{riechers11}.  
This suggests a short and vigorous burst of star formation in 
SDSS~J1607.  



\subsection{COSMOS-AzTEC1}

COSMOS-AzTEC1 is a source discovered by the AzTEC survey on 
JCMT \citep{scott08}, followed up by the Submillimeter Array \citep{younger07}.
Subsequent Keck DEIMOS spectroscopy has derived a redshift of 
$z \approx 4.64$ but the CO~(5--4) emission was not detected in the redshift 
range 4.56 -- 4.76 and 4.94 -- 5.02 \citep{smolcic11}.  We searched for 
the CO emission in the redshift range 4.38 -- 4.56, but 
did not detect any significant emission with a 3-sigma mass upper limit of 
$1.9 \times 10^{11}$~M$_{\odot}$, 
assuming a line width of 400~km~s$^{-1}$ and 
CO~(5--4) to CO~(1--0) ratio of 0.32 \citep{bothwell12}.  As discussed in \citet{smolcic11}, 
these non-detections could imply a low CO excitation 
or an incorrect redshift.


\section{Summary and Future Prospects}

We present new CO observations toward five submm bright high redshift
sources from the upgraded Nobeyama 45m telescope.  
All of the detected sources have large molecular gas mass 
(M$_{\rm H_2} \sim 10^{11}$~M$_{\odot}$) and  
line-widths (430 -- 530~km~s$^{-1}$).
One source (Orochi) shows a double peak spectrum.
We plan to expand this project by tuning the two different 
polarizations to different frequencies,
allowing us to instantaneously observe 16~GHz bandwidth in a 
single beam.  
Further, a new dual-color TES bolometer camera (Oshima et al., in prep)  
mounted on ASTE 
will detect and provide constraints to the redshifts of the SMGs.  
The bright (S$_{850} \gtrsim 10$~mJy) sources 
will be excellent targets for CO followup at the 45m.  

We thank our referee for useful comments.
We are grateful for the support by 
Fujitsu Limited, and Elecs Industry Co., Ltd.
This work is supported by Research Fellowship for Young Scientists from the
Japan Society of the Promotion of Science (JSPS) (BH),   
KAKENHI (no. 23840007) (YT), and 
the Grant-in-Aid for Specially Promoted Research from JSPS (no. 20001003).



\bigskip


\end{document}